\documentclass[superscriptaddress,
twocolumn,secnumarabic,nofootinbib]{revtex4-1}
\usepackage{amsmath}
\usepackage[utf8]{inputenc}
\usepackage[T1]{fontenc}

\pdfoutput=1

\usepackage{bm}
\usepackage{times}
\usepackage{braket}
\usepackage{color,graphicx}
\usepackage[utf8]{inputenc}
\usepackage[T1]{fontenc}
\usepackage[export]{adjustbox}

\usepackage{enumitem}

\usepackage{verbatim}

\usepackage{float}

\usepackage{colortbl}
\usepackage{longtable}
\usepackage{tabularx}
\usepackage{makecell}

\usepackage{multirow}

\usepackage{xcolor}

\usepackage{lipsum}

\usepackage[normalem]{ulem}

\definecolor{LightGray}{gray}{0.96}
\definecolor{Gray}{gray}{0.94}
\definecolor{nicered}{rgb}{0.7,0.1,0.1}
\definecolor{nicegreen}{rgb}{0.1,0.5,0.1}

\usepackage[colorlinks,
            linkcolor=black,
            filecolor=black,
            anchorcolor=black,
            urlcolor=black,
            citecolor=blue,
            bookmarks=false,
            ]{hyperref}

\allowdisplaybreaks

\setlongtables

\begin{document}

\title{Explaining $b\to s \ell^+ \ell^-$ data by sneutrinos in the $R$-parity violating MSSM}


\author{Quan-Yi Hu} \email[]{qyhu@aynu.edu.cn}
\author{Lin-Lin Huang} \email[]{huanglinlin@aynu.edu.cn}
\affiliation{School of Physics and Electrical Engineering, Anyang Normal University, Anyang, Henan 455000, China}
  
\begin{abstract}
The recent measurements on $b\to s \ell^+ \ell^-$ processes suggest the existence of lepton-flavour-universality breaking new physics. In this work, we have explored the possibility of explaining these data by sneutrinos in the $R$-parity violating minimal supersymmetric standard model. We study the light sneutrinos, of order 1 TeV, and suppose that the rest of the sfermions are much heavier than them. This setup can solve the $b \to s \mu^+ \mu^-$ anomaly well, and it is almost unconstrained by other related processes, such as $B_s-\bar{B}_s$ mixing, as well as $B_s^0 \to \tau^+ \tau^-$, $B^+ \to K^+ \tau^+ \tau^-$, $B_s^0 \to \tau^\pm \mu^\mp$, $B^+ \to K^+ \tau^\pm \mu^\mp$, and $B \to K^{(\ast)} \nu \bar{\nu}$ decays. 
\end{abstract}
\maketitle

\section{Introduction}
\label{sec:introduction}

The rare semileptonic $b$-hadron decays induced by the flavour-changing neutral current (FCNC) transition $b\to s \ell^+ \ell^-$ do not arise at tree level and are highly suppressed at higher orders within the Standard Model (SM), due to the Glashow-Iliopoulos-Maiani (GIM) mechanism~\cite{Glashow:1970gm}. New TeV-scale particles in many extensions of the SM could lead to measurable effects in these rare decays. As a consequence, they play a crucial role in testing the SM and probing various new physics (NP) scenarios beyond it~\cite{Hurth:2010tk,Blake:2016olu}.

In recent years, several deviations from the SM predictions have been observed in $b\to s \ell^+ \ell^-$ transition. Consider the ratios of the branching fractions $R_{K^{(\ast)}} = {\cal B}(B \to K^{(\ast)} \mu^+ \mu^-) / {\cal B}(B \to K^{(\ast)} e^+ e^-)$, which have negligible theoretical uncertainties. In the range $1.1 < q^2 < 6 {\rm GeV}^2/c^4$, the latest experimental data by LHCb collaboration give $R_K^{ [1.1, 6]} = 0.846^{+0.060 +0.016}_{-0.054-0.014}$~\cite{Aaij:2019wad,Aaij:2014ora}, but the SM predicts it to be close to one~\cite{Bordone:2016gaq}. The measurement of $R_K$ is $2.5\sigma$ smaller than the SM prediction. The measurements of $R_{K^\ast}$~\cite{Aaij:2017vbb} by LHCb are $R_{K^\ast}^{[0.045, 1.1]} = 0.66^{+0.11}_{-0.07} \pm 0.03$ and $R_{K^\ast}^{[1.1, 6.0]} = 0.69^{+0.11}_{-0.07} \pm 0.05$, which are lower than the predicted values of the SM~\cite{Bordone:2016gaq} about $2.1 \sigma$ and $2.5 \sigma$, respectively. Belle collaboration also give the measurements of $R_{K^{(\ast)}}$~\cite{Abdesselam:2019lab,Abdesselam:2019wac}, which are consistent with the SM predictions due to their large experimental errors. In addition to the tension with the SM in lepton-flavour-universality observables $R_{K^{(\ast)}}$, some other deviations have also been found in $b \to s \mu^+ \mu^-$ transition. In particular, the form-factor-independent angular observable $P_5'$~\cite{DescotesGenon:2012zf,Descotes-Genon:2013vna,Hu:2016gpe} in the $B \to K^\ast \mu^+ \mu^-$ decay was measured by LHCb~\cite{Aaij:2015oid,Aaij:2013qta}, CMS~\cite{Khachatryan:2015isa}, ATLAS~\cite{Aaboud:2018krd} and Belle~\cite{Wehle:2016yoi,Abdesselam:2016llu}, showing a $2.6 \sigma$ disagreement with the SM expectation~\cite{Alguero:2019ptt}. Finally, LHCb has also observed a $3.3\sigma$ deficit in the $B_s^0 \to \phi \mu^+ \mu^-$ decay~\cite{Aaij:2015esa,Aaij:2013aln}. 

Motivated by these deviations and using the other available data on such rare $b\to s \ell^+ \ell^-$ transitions, many global analyses have been carried out~\cite{Aebischer:2019mlg,Alok:2019ufo,Alguero:2019ptt,Ciuchini:2019usw,Arbey:2019duh,Kowalska:2019ley,Capdevila:2019tsi,Bhattacharya:2019dot}, finding that a negative shift in a single Wilson coefficient of local operator like $O_9^{\mu\mu} = ({\bar s} \gamma^\alpha P_L b)({\bar \mu} \gamma_\alpha \mu)$ or $O_{LL}^{\mu\mu} = ({\bar s} \gamma^\alpha P_L b)({\bar \mu} \gamma_\alpha P_L \mu)$ leads to a consistent description of the data, with the corresponding best-fit point can improve the fit to the data by more than $5 \sigma$ compared to the SM. Furthermore, the operator $O_{LL}^{\mu\mu}$ performs better than $O_9^{\mu\mu}$, mainly because there is now $\sim 2\sigma$ tension in the branching fraction of $B_s \to \mu^+ \mu^-$~\cite{Aaij:2013aka,Chatrchyan:2013bka,CMS:2014xfa,Aaboud:2016ire,Aaij:2017vad,Aaboud:2018mst,Aebischer:2019mlg}, which is not affected by $O_9^{\mu\mu}$. In this paper, we work with the low-energy effective weak Lagrangian governing the $b \to s \mu^+ \mu^-$ processes:
\begin{equation}
{\cal L}_{\rm eff} = {\cal L}_{\rm eff}^{\rm SM} + \frac{4 G_F}{\sqrt{2}} \eta_t \frac{e^2}{16 \pi^2} C_{LL}^{\mu \mu} O_{LL}^{\mu\mu} +{\rm H.c.}\,,
\end{equation} 
where ${\cal L}_{\rm eff}^{\rm SM}$ represents contributions from the SM, and the remaining terms contain possible NP contributions. The CKM factor $\eta_t = V_{tb}V_{ts}^\ast \approx -0.04$~\cite{Tanabashi:2018oca}. The best-fit point performed by Ref.~\cite{Aebischer:2019mlg} is $C_{LL}^{\mu \mu} = -1.06$, with the $2\sigma$ range being $-1.38 < C_{LL}^{\mu \mu} < -0.74$. We find that such $C_{LL}^{\mu \mu}$ can be generated naturally in the $R$-parity violating minimal supersymmetric standard model (MSSM)~\cite{Barbier:2004ez} by exchanging muon sneutrinos and winos. 

Before we start our discussion, let us briefly review some of the work on $b\to s \mu^+ \mu^-$ anomaly within the context of $R$-parity violating MSSM~\cite{Biswas:2014gga,Deshpand:2016cpw,Das:2017kfo,Earl:2018snx,Darme:2018hqg,Trifinopoulos:2018rna,Trifinopoulos:2019lyo}. For example, the authors in Ref.~\cite{Deshpand:2016cpw} attempt to explain $b \to s \mu^+ \mu^-$ anomaly via one-loop contributions involving right-handed down type squarks $\tilde{d}_{R}$, which can help solve $R(D^{\ast})$ anomaly at tree level~\cite{Deshpande:2012rr,Zhu:2016xdg,Deshpand:2016cpw,Altmannshofer:2017poe,Trifinopoulos:2018rna,Hu:2018lmk,Hu:2018veh}. However, they note that it is difficult to find a viable explanation due to the severe constraints from the upper limit on the branching fraction of $B \to K^{(\ast)} \nu \bar{\nu}$ decays. In addition to $\tilde{d}_{R}$, the authors in Ref.~\cite{Das:2017kfo} also consider the contribution to $b\to s \mu^+ \mu^-$ transition from the box diagrams with a left-handed up type squark $\tilde{u}_{L}$ and sneutrino $\tilde{\nu}_{L}$ in the loop. They find that this new contribution could help explain $b \to s \mu^+ \mu^-$ anomaly, while satisfying the constraint from $B \to K^{(\ast)} \nu \bar{\nu}$ and $D^0 \to \mu^+ \mu^-$ decays as well as $B_s - \bar B_s$ mixing. In Ref.~\cite{Earl:2018snx}, the authors focus on parameters for which diagrams involving winos $\tilde{W}$, which have not been considered before, make significant effects. They set the masses of $\tilde{W}$ and three $\tilde{u}_{L}$ to be light, of order 1 TeV, and at the same time, they consider heavy $\tilde{\nu}_{L}$ and  $\tilde{d}_{R}$, of order 10 TeV. In this scenario, the $b \to s \mu^+ \mu^-$ anomaly may be explained by large values of $\lambda'$, but the available parameter space is very small due to the constraints from relevant processes, such as $\tau \to 3\mu$, $B_s - \bar B_s$ mixing and direct LHC searches. The restriction from $B \to K^{(\ast)} \nu \bar{\nu}$ decay is negligible because of the large mass of $\tilde{d}_{R}$. 

There are two kinds of sfermions participating in the $\tilde{W}$ box diagrams, namely $\tilde{u}_{L}$ and $\tilde{\nu}_{L}$. As an alternative, in this paper, we study the light $\tilde{\nu}_{L}$, of order 1 TeV, and suppose that the rest of sfermions are much heavier (a few 10 TeV or larger) compared to it. This scenario can well produce the $C_{LL}^{\mu \mu}$ needed to explain $b\to s \mu^+ \mu^-$ anomaly, and the corresponding parameter space is not constrained by other related processes, such as $B_s-\bar{B}_s$ mixing, as well as $B_s^0 \to \tau^+ \tau^-$, $B^+ \to K^+ \tau^+ \tau^-$, $B_s^0 \to \tau^\pm \mu^\mp$, $B^+ \to K^+ \tau^\pm \mu^\mp$ and $B \to K^{(\ast)} \nu \bar{\nu}$ decays.  

Our paper is organized as follows. In Sec.~\ref{sec:bsuu_RPVMSSM}, we first set up our scenario and then discuss the explanation of $b \to s \mu^+ \mu^-$ anomaly in the $R$-parity violating MSSM. The other potential constraints are studied in Sec.~\ref{sec:constraints}. Our conclusions are finally made in Sec.~\ref{sec:conclusion}.

\section{Contributions to $b \to s \mu^+ \mu^-$ processes from $R$-parity violating MSSM}
\label{sec:bsuu_RPVMSSM}

\begin{figure}[t]
	\centering
	\includegraphics[width=0.42\textwidth,pagebox=cropbox,clip]{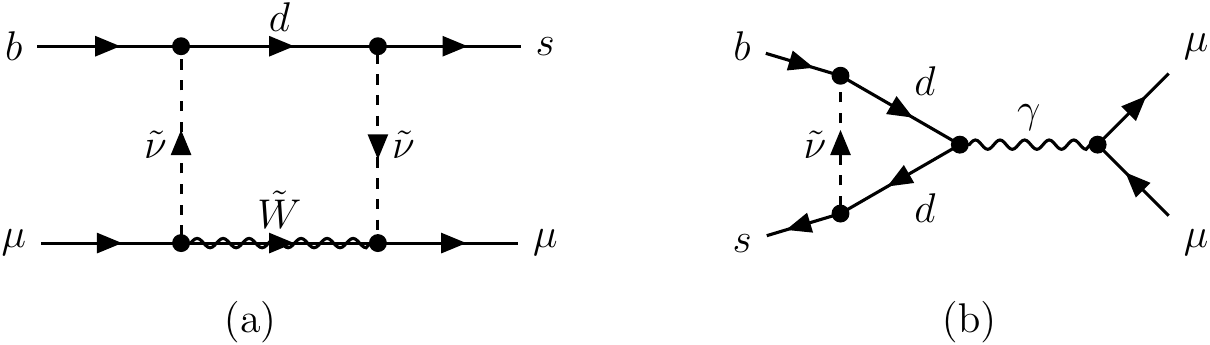}
	\caption{\label{fig:feynman} \small Feynman diagrams for $b \to s \mu^+ \mu^-$ transition in our scenario.}
\end{figure}

The superpotential of the relevant $R$-parity violating terms in the MSSM is given by~\cite{Barbier:2004ez}
\begin{align}
\label{eq:W_RPV}
W_{\rm RPV} = &\, \mu_i L_i H_u + \frac{1}{2} \lambda_{ijk} L_i L_j E_k^c + \lambda'_{ijk} L_i Q_j D_k^c\nonumber \\[2mm]
 &+ \frac{1}{2} \lambda''_{ijk} U_i^c D_j^c D_k^c\, ,
\end{align}
where $L$, $H_u$, $E^c$, $Q$, $D^c$, and $U^c$ are the chiral superfields for the MSSM multiplet, and we denote the generation indices by $i,j,k = 1,2,3$. The summation is applied for the repeated indices throughout this paper unless otherwise stated. The first three terms in Eq.~\eqref{eq:W_RPV} destroy the lepton number and the last term violates the baryon number. We will assume that $\lambda''$ coupling is zero to prevent rapid proton decay. In this work, we limit ourselves to consider the $\lambda'_{ijk} L_i Q_j D_k^c$ term as the source of $R$-parity violating NP, because of the $b \to s \mu^+ \mu^-$ processes involve both leptons and quarks. The effects of $\lambda$ and $\lambda'$ terms simultaneously on $b \to s \mu^+ \mu^-$ processes have been studied in Refs.~\cite{Trifinopoulos:2018rna,Trifinopoulos:2019lyo}. Expanding the chiral superfields in terms of their fermions and sfermions, one has
\begin{align}
\label{eq:calL}
{\cal L} = &\, \lambda'_{ijk}\big(\tilde{\nu}_{Li} \bar{d}_{Rk} d_{Lj} + \tilde{d}_{Lj} \bar{d}_{Rk} \nu_{Li} + \tilde{d}_{Rk}^\ast \bar{\nu}_{Li}^c d_{Lj}\nonumber \\[2mm] 
&- \tilde{l}_{Li} \bar{d}_{Rk} u_{Lj} - \tilde{u}_{Lj} \bar{d}_{Rk} l_{Li} - \tilde{d}_{Rk}^\ast \bar{l}_{Li}^c u_{Lj}\big)\,.
\end{align}
We assume that all sfermions are so heavy (a few 10 TeV or larger) that they are decoupled\footnote{That is to say, the Feynman diagram that contains these heavy sfermions does not have to be considered because their contributions are suppressed by $\mu^2_{\rm EW}/m^2_{\tilde \psi} \sim 10^{-4}$.}, except sneutrinos $\tilde{\nu}_{Li}$ of order 1 TeV. Under this assumption\footnote{Due to the $SU(2)_L$ symmetry, the left-handed charged sleptons
may have a mass comparable to that of the sneutrinos and can affect $b \to s \mu^+ \mu^-$ processes by exchanging smuon and neutralino. The Feynman diagram is similar to the box diagram where the two scalar lines in Fig.~\ref{fig:feynman}a crossing and has negligible effects due to the assumption $\lambda'_{ij2} =0$, so the discussion in this work does not include the charged sleptons.}, only the $\lambda'_{ijk} \tilde{\nu}_{Li} \bar{d}_{Rk} d_{Lj}$ term in Eq.~\eqref{eq:calL} can lead to a valuable effect. These interactions are similar to but different from the generic terms $\bar \Psi_A (L^b_{AM} P_L b + L^s_{AM} P_L s + L^\mu_{AM} P_L \mu) \Phi_M$ given in Ref.~\cite{Arnan:2019uhr}. In our work the fermions $\Psi_A$ represent the SM down type quarks rather than new particles, and the interactions with charged leptons are provided by $R$-parity conserving MSSM. In this paper we focus our attention on a parameter space where the $\lambda'_{ij3}$ couplings are large, i.e., keep $\lambda'_{ij1} = \lambda'_{ij2} =0$ all the time. We will assume sneutrinos are in their mass eigenstate basis and nearly degenerate, and the degenerate mass is denoted as $m_{\tilde{\nu}}$. We should further assume that $\lambda'_{i33} \lambda'^\ast_{i23} = 0$, which discards NP contributions to all related channels without explicit external leptons, as exploited in the following $b s \gamma$-vertex of photonic penguin and in Sec.~\ref{sec:constraints}.

\begin{figure}[t]
	\centering
	\includegraphics[width=0.4\textwidth,pagebox=cropbox,clip]{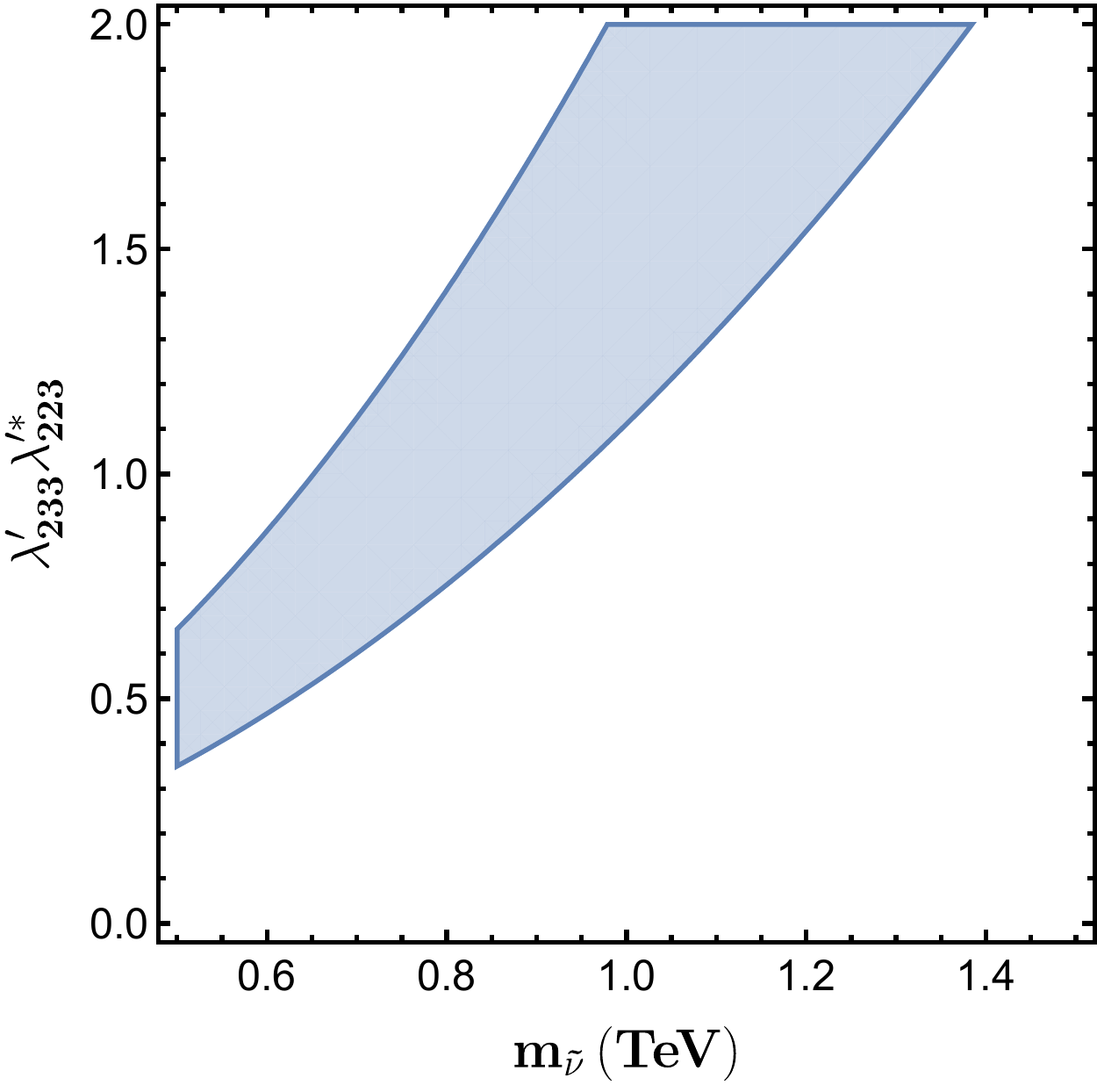}
	\caption{\label{fig:result} \small The figure showing the parameter space in $m_{\tilde{\nu}}-\lambda'_{233}\lambda'^\ast_{223}$ plane explaining $b \to s \mu^+ \mu^-$ anomaly. We set $m_{\tilde{W}} = 0.3$~TeV.}
\end{figure}

The $b \to s \mu^+ \mu^-$ processes can occur at one-loop level by exchanging muon sneutrinos and winos, see Fig.~\ref{fig:feynman}a (The box diagram where the two sneutrino lines crossing is discarded due to $\lambda'_{ij2} =0$). After integrating out the sparticles we are left with the effective operator $O_{LL}^{\mu\mu}$, as well as the corresponding Wilson coefficient given by
\begin{equation}
\label{eq:CLLmu}
C_{LL}^{\mu\mu} = -\frac{\sqrt{2} \lambda'_{233}\lambda'^\ast_{223}}{16 G_F \sin^2\theta_W \eta_t m^2_{\tilde{\nu}}}  x_{\tilde{\nu}} f(x_{\tilde{\nu}})\, ,
\end{equation}
where the loop function $f(x_{\tilde{\nu}}) \equiv \frac{ 1-x_{\tilde{\nu}}+\log{x_{\tilde{\nu}}} }{(1-x_{\tilde{\nu}})^2}$ and $x_{\tilde{\nu}} \equiv m^2_{\tilde{\nu}}/m^2_{\tilde{W}}$. To explain $b\to s \mu^+ \mu^-$ anomaly, we need to take the product $\lambda'_{233}\lambda'^\ast_{223} > 0$ to make $C_{LL}^{\mu\mu}$ negative. Consider the $2\sigma$ range $-1.38 < C_{LL}^{\mu \mu} < -0.74$~\cite{Aebischer:2019mlg}, we have
\begin{equation}
\label{eq:mainres}
-1.74 < \frac{x_{\tilde{\nu}} f(x_{\tilde{\nu}}) \lambda'_{233}\lambda'^\ast_{223}}{(m_{\tilde{\nu}}/{\rm TeV})^2} < -0.93\, .
\end{equation}
The corresponding parameter space is shown in Fig.~\ref{fig:result}.

There is also a contribution from the photonic penguin, which is shown in Fig.~\ref{fig:feynman}b. In fact, this contribution is lepton flavour universal because of the SM photon. Using {\tt FeynCalc}~\cite{Mertig:1990an,Shtabovenko:2016sxi} and {\tt Package-X}~\cite{Patel:2015tea,Patel:2016fam} packages, we can obtain the effective operators $O_9^{\ell\ell}$ and $O_{7}=\frac{m_b}{e}(\bar s \sigma^{\alpha \beta} P_R b) F_{\alpha \beta}$ after integrating out sneutrinos, and the corresponding Wilson coefficients given by 
\begin{align}
C_9^{\ell \ell} &= -\frac{\sqrt{2}\lambda'_{i33} \lambda'^\ast_{i23}}{36 G_F \eta_t m^2_{\tilde{\nu}}}\left[\frac{4}{3} + \log\left(\frac{m_b^2}{m^2_{\tilde{\nu}}}\right)\right]\, ,
\label{eq:C9ll}\\[2mm]
C_7 &= \frac{\sqrt{2}\lambda'_{i33} \lambda'^\ast_{i23}}{144 G_F \eta_t m^2_{\tilde{\nu}}} \, .
\label{eq:C7gamma}
\end{align}
Our results are consistent with those in Ref.~\cite{deGouvea:2000cf}. Comparing with Ref.~\cite{Earl:2018snx}, we find that the result of $C_7$ is consistent, but the result of $C_9^{\ell \ell}$ is different by a negative sign. All in all, we should remove the effect of photonic penguin by assuming $\lambda'_{i33} \lambda'^\ast_{i23} = 0$ in order to take advantage of only nonzero $C_{LL}^{\mu \mu}$ scenario, which has the largest pull-value in single Wilson coefficient global analyses~\cite{Aebischer:2019mlg}. Similarly, there is no contribution of $Z$-penguin under the assumption $\lambda'_{i33} \lambda'^\ast_{i23} = 0$. 

\section{Other possible constraints}
\label{sec:constraints}

In our scenario, several other processes may also obtain the effects of $R$-parity violating interactions, and the corresponding constraints should be taken into account. Next, we mainly study the constraints on $\lambda'_{i23}$ and $\lambda'_{i33}$ couplings, which play the key role in solving $b\to s \mu^+ \mu^-$ anomaly.

\subsection{Tree level decays}
\label{subsec:tree}

Exchanging sneutrinos and performing Fierz rearrangement, one obtains the following four fermion operators at tree level
\begin{equation}
\label{eq:Lefftree}
{\cal L}_{\rm eff}^{\rm tree} = \frac{\lambda'_{ij3} \lambda'^\ast_{ij'3}}{ m^2_{\tilde{\nu}}} \left(\bar{b}_{R} d_j\right) \left(\bar{d}_{j'} b_R\right)\,.
\end{equation}
There is no valid constraint here. In addition, assuming $\lambda'_{i33} \lambda'^\ast_{i23} = 0$ can prevent the occurrence of dangerous $\Upsilon-B_s$ mixing.

\subsection{Loop level decays}
\label{subsec:loop}

The potential constraint may come from $B_s-\bar{B}_s$ mixing, which is induced by one loop diagrams. In our scenario, this constraint vanishes due to the assumption $\lambda'_{i33} \lambda'^\ast_{i23} = 0$. 

In fact, in addition to the muon channel, the nonzero $\lambda'_{i23}$ and $\lambda'_{i33}$ couplings can also induce $b\to s \ell^+_{i} \ell^-_{j}$ processes by exchanging sneutrinos and winos, as shown in Fig.~\ref{fig:feynman}a. The corresponding Wilson coefficients $C_{LL}^{ij}$ can be obtained by replacing $\lambda'_{233}\lambda'^\ast_{223}$ with $\lambda'_{i33}\lambda'^\ast_{j23}$ in Eq.~\eqref{eq:CLLmu}. In order for the NP to have no effect on $b \to s e^+ e^-$ processes we should keep $C_{LL}^{ee} =\lambda'_{133}\lambda'^\ast_{123} \approx 0$, which means $\lambda'_{133} \approx 0$ or $\lambda'^\ast_{123} \approx 0$. Combining $\lambda'_{i33} \lambda'^\ast_{i23} = 0$, we predict the same size of $C_{LL}^{\mu \mu}$ and $C_{LL}^{\tau \tau} \propto \lambda'_{333}\lambda'^\ast_{323}$, with similar result in the ${\rm PS}^3$ model~\cite{Bordone:2018nbg}. Such $C_{LL}^{\tau \tau}$ satisfies the upper limit of ${\cal B} (B^+ \to K^+ \tau^+ \tau^-) < 2.25 \times 10^{-3}$~\cite{TheBaBar:2016xwe} measured by BaBar at 90\% confidence level (CL) and ${\cal B}(B_s^0 \to \tau^+ \tau^-) < 6.8 \times 10^{-3}$~\cite{Aaij:2017xqt} measured by LHCb at 95\% CL.

The remaining potential constraints come from several lepton-flavour-violation decays $B_s^0 \to \tau^\pm \mu^\mp$ and $B^+ \to K^+ \tau^\pm \mu^\mp$. Those decays governed by the low-energy effective weak Lagrangian
\begin{equation}
\label{eq:Lbsij}
{\cal L}_{\rm eff}^{b\to s \ell_i^+ \ell_j^-} = - \frac{\alpha}{16 \pi \sin^2\theta_W} \frac{\lambda'_{i33}\lambda'^\ast_{j23}}{m^2_{\tilde{\nu}}} x_{\tilde{\nu}} f(x_{\tilde{\nu}}) O_{LL}^{ij} +{\rm H.c.}\,,
\end{equation}
with $i\neq j$. The branching fractions of leptonic $B_s^0 \to \tau^\pm \mu^\mp$ decays given by~\cite{Becirevic:2016zri}
\begin{align}
{\cal B}(B_s^0 \to \tau^+ \mu^- ) = \frac{\alpha^2 \tau_{B_s} f_{B_s}^2 \lambda_{\tau \mu} x^2_{\tilde{\nu}} f^2(x_{\tilde{\nu}})}{128^2 \pi^3  m_{B_s}^3 \sin^4\theta_W}  \frac{|\lambda'_{333}\lambda'^\ast_{223}|^2}{m^4_{\tilde{\nu}}}\, , \\[2mm]
{\cal B}(B_s^0 \to \mu^+ \tau^-) = \frac{\alpha^2 \tau_{B_s} f_{B_s}^2 \lambda_{\tau \mu} x^2_{\tilde{\nu}} f^2(x_{\tilde{\nu}})}{128^2 \pi^3  m_{B_s}^3 \sin^4\theta_W}  \frac{|\lambda'_{233}\lambda'^\ast_{323}|^2}{m^4_{\tilde{\nu}}}\, ,
\end{align}
where
\begin{align}
\lambda_{\tau \mu} \equiv &
\left[m_{B_s}^2( m_{\tau}^2 + m_{\mu}^2) - ( m_{\tau}^2 -  m_{\mu}^2)^2\right]\nonumber \\[2mm]
&\times \sqrt{\left(m^2_{B_s} - m^2_{\tau} - m^2_{\mu}\right)^2 - 4m^2_{\tau}m^2_{\mu}}.
\end{align}
In our numerical analysis, we take as input the decay constant $f_{B_s} = 0.2272(34)$~GeV, the lifetime $\tau_{B_s} = 1.510(4)$~ps, as well as the mass $m_{B_s} =5.367$~GeV, $m_{\tau} = 1.777$~GeV and  $m_{\mu} = 0.1057$~GeV~\cite{Tanabashi:2018oca}. Lately, the upper limit on these branching fractions are measured by LHCb collaboration. At 95\% CL one has~\cite{Aaij:2019okb}
\begin{equation}
\label{eq:Bsexp}
{\cal B}(B_s^0 \to \tau^\pm \mu^\pm)_{\rm exp} < 4.2\times 10^{-5}\, .
\end{equation}
This induces the constraints
\begin{equation}
\label{eq:lepres}
\frac{|x_{\tilde{\nu}} f(x_{\tilde{\nu}}) \lambda'_{333}\lambda'^\ast_{223} (\lambda'_{233}\lambda'^\ast_{323})|}{(m_{\tilde{\nu}}/{\rm TeV})^2} < 108.15\, .
\end{equation}
For semileptonic $B^+ \to K^+ \tau^\pm \mu^\mp$ decays, we can obtain
\begin{align}
\frac{|x_{\tilde{\nu}} f(x_{\tilde{\nu}}) \lambda'_{333}\lambda'^\ast_{223}|}{(m_{\tilde{\nu}}/{\rm TeV})^2} &< 92.24\, ,\label{eq:semilepres1} \\[2mm]
\frac{|x_{\tilde{\nu}} f(x_{\tilde{\nu}}) \lambda'_{233}\lambda'^\ast_{323}|}{(m_{\tilde{\nu}}/{\rm TeV})^2} &< 117.53\, ,\label{eq:semilepres2}
\end{align}
by directly using the upper bound results of the Wilson coefficients given in Ref.~\cite{Barbieri:2019zdz}. The Eq.~\eqref{eq:semilepres1} has a stronger constraint than Eq.~\eqref{eq:lepres} but Eq.~\eqref{eq:semilepres2} has a weaker constraint than it. Obviously, under these constraints the Eq.~\eqref{eq:mainres} and relation $\lambda'_{233}\lambda'^\ast_{223} \approx -\lambda'_{333}\lambda'^\ast_{323}$ (for keeping $\lambda'_{i33} \lambda'^\ast_{i23} = 0$) are easy to implement.

Finally, we discuss the influence of sneutrinos on $b \to s \bar\nu_i \nu_j$ processes, which can obtain the $R$-parity violating contributions by exchanging sneutrinos and neutralinos in the loop. This NP contribution can lead to the same effective operator as the SM.  The effective Lagrangian for these processes are defined by 
\begin{equation}
\label{eq:bsnunuleff}
{\cal L}_{\rm eff} = (C_{LL}^{\rm SM} \delta_{ij} + C_{LL}^{\bar{\nu}_i \nu_j}) ({\bar s} \gamma^\alpha P_L b)({\bar \nu}_j \gamma_\alpha P_L \nu_i) + {\rm H.c.}\, ,
\end{equation}
where~\cite{Buras:2014fpa}
\begin{equation}
C_{LL}^{\rm SM} = - \frac{\sqrt{2}G_F \alpha \eta_t X_t}{\pi \sin^2\theta_W},\; X_t = 1.469 \pm 0.017\, , 
\end{equation}
 is generated by the SM. The contributions of $R$-parity violating interactions are given by
 \begin{equation}
C_{LL}^{\bar{\nu}_i \nu_j} = - \frac{\alpha}{32 \pi } \frac{\lambda'_{i33}\lambda'^\ast_{j23}}{m^2_{\tilde{\nu}}}\left[\frac{x_{\tilde{\nu}} f(x_{\tilde{\nu}})}{\sin^2\theta_W} + \frac{y_{\tilde{\nu}} f(y_{\tilde{\nu}})}{\cos^2\theta_W}\right]\, ,
\end{equation}
where $y_{\tilde{\nu}} \equiv m^2_{\tilde{\nu}}/m^2_{\tilde{B}}$, $m_{\tilde{B}}$ is the bino mass. It is useful to define the ratio $R_{B \to K^{(\ast)}\nu\nu} \equiv {\cal B}(B \to K^{(\ast)}\nu\nu) / {\cal B}(B \to K^{(\ast)}\nu\nu)_{\rm SM} $, and it is given by
\begin{align}
\label{eq:Rbsnnu}
&R_{B \to K^{(\ast)}\nu\nu} = \frac{\sum\limits_{i=1}^3 \left| C_{LL}^{\rm SM} + C_{LL}^{\bar{\nu}_i \nu_i}\right|^2 + \sum\limits_{i \neq j}^3 \left| C_{LL}^{\bar{\nu}_i \nu_j} \right|^2}{3 \left| C_{LL}^{\rm SM} \right|^2}\nonumber \\[2mm] 
&\quad = 1 + \frac{2 \left|  C_{LL}^{\bar{\nu}_2 \nu_2} \right|^2 +\left|  C_{LL}^{\bar{\nu}_2 \nu_3} \right|^2 + \left|  C_{LL}^{\bar{\nu}_3 \nu_2} \right|^2}{3 \left| C_{LL}^{\rm SM} \right|^2}.
\end{align}
Because of $\lambda'_{233}\lambda'^\ast_{223} \approx -\lambda'_{333}\lambda'^\ast_{323}$, the interference term between the NP and the SM disappears. Let $m_{\tilde{B}} = m_{\tilde{W}}$ we have
\begin{align}
\label{eq:Rbsnunures}
&R_{B \to K^{(\ast)}\nu\nu} =1 + 5.9\times 10^{-4} \frac{x^2_{\tilde{\nu}} f^2(x_{\tilde{\nu}}) }{(m_{\tilde{\nu}}/{\rm TeV})^4}\nonumber \\[2mm]
&\quad \times \left( 2\left| \lambda'_{233}\lambda'^\ast_{223} \right|^2 + \left| \lambda'_{233}\lambda'^\ast_{323} \right|^2 + \left| \lambda'_{333}\lambda'^\ast_{223} \right|^2 \right) .
\end{align}
When the parameters fall into the interval given in Eq.~\eqref{eq:mainres}, the Eq.~\eqref{eq:Rbsnunures} satisfies the constraint from upper bounds $R_{B \to K \nu\nu} < 3.9$ and $R_{B \to K^\ast \nu\nu} < 2.7$~\cite{Grygier:2017tzo}, which are measured by Belle at 90\% CL.

\section{Conclusions}
\label{sec:conclusion}

Recently, several deviations from the SM predictions in $b\to s \ell^+ \ell^-$ data suggest the existence of NP which breaks the lepton-flavour universality. Many global analyses show that a negative shift in Wilson coefficient $C_{LL}^{\mu \mu}$ can explain these data well, and the corresponding best-fit point can improve the fit to the data by more than $5 \sigma$ compared to the SM. This suggests that the NP primarily affects the $b\to s \mu^+ \mu^-$ processes. Based on these knowledge, in this work we have explored the possibility of explaining $b\to s \mu^+ \mu^-$ anomaly by sneutrinos in the $R$-parity violating MSSM.

After a brief introduction to the relevant terms in the superpotential of $R$-parity violating MSSM, we present our scenario, that is, we consider the light $\tilde{\nu}_{L}$ of order 1 TeV and the other sfermions are so heavy (a few 10 TeV or larger) that they are decoupled. We find that a positive product $\lambda'_{233}\lambda'^\ast_{223}$ can explain $b\to s \mu^+ \mu^-$ anomaly, and the parameter space satisfied by $\lambda'_{233}\lambda'^\ast_{223}$ and $m_{\tilde{\nu}}$ is shown in Fig.~\ref{fig:result}. After that, we consider the other possible constraints, including tree level and one-loop level decays. Assuming $\lambda'_{i33} \lambda'^\ast_{i23} = 0$ can inhibit the contribution of $R$-parity violating NP to $B_s-\bar{B}_s$ mixing and the photonic penguin of $b\to s \ell^+ \ell^-$ processes, and prevents the emergence of dangerous $\Upsilon-B_s$ mixing. We predict $C_{LL}^{\tau \tau} \approx -C_{LL}^{\mu \mu}$ which satisfies the upper limit of the branching fractions of $B^+ \to K^+ \tau^+ \tau^-$ and $B_s^0 \to \tau^+ \tau^-$ decays. Furthermore, we discuss the potential constraints from $B_s^0 \to \tau^\pm \mu^\mp$, $B^+ \to K^+ \tau^\pm \mu^\mp$ and $B \to K^{(\ast)} \nu \bar{\nu}$ decays, and find that the experimental upper limit of these processes do not effectively exclude the parameter space needed to explain $b\to s \mu^+ \mu^-$ anomaly.

\section*{Acknowledgements}

The authors thank Min-Di Zheng for useful communications. This work is supported by the National Natural Science Foundation of China under Grant Nos.~11947083.


\end{document}